\documentclass[12pt]{spieman}
\usepackage{amsmath,amssymb,amsfonts}
\usepackage{graphicx}
\usepackage{booktabs}
\usepackage{array}
\usepackage{setspace}
\usepackage{tocloft}
\usepackage{lineno}
\usepackage{url}
\newcommand{\doi}[1]{\href{https://doi.org/#1}{doi:\nolinkurl{#1}}}
\usepackage{placeins}
\usepackage{flafter}
\title{Spectral-spatial vision test-time training for hyperspectral image classification}
\author[a,*]{Xiang Zhao}
\affil[a]{College of Mathematical Sciences, Harbin Engineering University, 145 Nantong Street, Nangang District, Harbin 150001, China}

\cftpagenumbersoff{figure}
\cftpagenumbersoff{table}
\begin{document}
\maketitle

\begin{abstract}
Hyperspectral image (HSI) classification requires the representation of contiguous spectral signatures and spatial context under limited labeled supervision. Vision test-time training (TTT) offers an alternative to fixed-weight token mixing by performing self-supervised inner-model updates on the input sequence, but its application to center-pixel HSI classification remains insufficiently characterized. We present ViTTT-HSI, a spectral--spatial adaptation of the vision test-time training (ViT$^3$) architecture. A spectral-aware embedding stem extracts ordered-band information using spectral convolution and forms spatial tokens without reducing their spatial resolution. Two ViT$^3$-based TTT encoder blocks perform patch-conditioned token mixing, followed by the fusion of center-pixel and neighborhood-average features for classification. We evaluate the framework on Indian Pines, Pavia University, Salinas, and Houston 2013 using ten random seeds and dataset-specific training fractions of 1\% or 5\%. The mean overall accuracies are 97.6390\%, 98.8753\%, 99.3292\%, and 98.4283\%, respectively. Ablations reveal a substantial contribution from the TTT inner update on Indian Pines and a smaller positive contribution on Pavia University, while the relative benefits of the spectral stem and feature fusion depend on the dataset. The results demonstrate the feasibility of using patch-conditioned vision TTT for spectral--spatial HSI classification and highlight the role of the inner adaptation mechanism.
\end{abstract}

{\noindent\footnotesize\textbf{*}Corresponding author: Xiang Zhao, \linkable{2021241331@hrbeu.edu.cn}}

\begin{spacing}{1}

\section{Introduction}
\label{sec:intro}
Hyperspectral images (HSIs) record a dense sequence of spectral measurements for each spatial location, providing discriminative signatures for identifying vegetation, built surfaces, and other land-cover materials. Their classification is a fundamental step in remote sensing applications such as agricultural monitoring, environmental assessment, and urban mapping. Compared with conventional three-channel images, HSI cubes contain substantially more spectral channels and often exhibit pronounced correlations between neighboring bands. At the same time, reliable pixel annotations can be costly to obtain, making accurate classification under limited supervision a persistent research challenge~\cite{paoletti2019review,jia2021fewshotsurvey}.

A classifier based solely on the spectrum of a target pixel may be sensitive to local spectral variation and similarities between distinct materials. Incorporating neighboring pixels provides complementary spatial evidence, but also requires the model to accommodate boundaries and heterogeneous patches. Deep convolutional networks have addressed this issue by learning local spectral and spatial features through 3-D and 2-D operators~\cite{zhong2018ssrn,roy2020hybridsn}. More recently, Transformers have offered flexible sequence-level interactions for spectral and spatial feature learning, as illustrated by SpectralFormer and the spectral--spatial feature tokenization Transformer (SSFTT)~\cite{hong2022spectralformer,sun2022ssftt}. Selective state-space models have also been used for hyperspectral representation learning, with S$^2$Mamba combining spatial and spectral sequence processing~\cite{wang2025s2mamba}.

These advances offer increasingly expressive spectral--spatial encoders. A complementary question is whether the token-mixing operation itself can adapt to the particular neighborhood being classified. Test-time training (TTT) addresses sequence modeling through temporary, input-dependent inner-model updates: features from the current sequence form key--value training pairs, the inner model is updated by an unlabeled objective, and the adapted model then processes query features~\cite{sun2025ttt}. In vision, Han et al. developed ViT$^3$ and studied how inner-training objectives, optimization settings, gated units, and depthwise convolutional inner models affect visual token mixing~\cite{han2026vit3}. This approach provides a patch-conditioned representation without requiring class labels at inference.

Applying a visual TTT architecture to HSI center-pixel recognition requires changes at the input and prediction interfaces. First, the spectral axis carries a meaningful band ordering that is absent from standard red, green, and blue (RGB) patch embeddings. Second, the target is the label of the center pixel rather than a single label for the entire image. Third, spatial positions should remain explicitly associated with their original neighborhoods so that center information can be recovered after token interaction. These requirements motivate a spectral-aware embedding and a center-oriented prediction head around a ViT$^3$-based spatial encoder.

In this work, we construct ViTTT-HSI, a spectral--spatial adaptation of vision TTT for HSI center-pixel classification. A spectral convolutional stem first extracts ordered-band responses and creates one token per spatial position. Two encoder blocks then apply patch-conditioned TTT mixing while retaining the complete spatial token grid. Finally, the normalized center token is combined with an average representation of the neighborhood for the class prediction. The TTT inner models and their update rules follow ViT$^3$~\cite{han2026vit3}; the hyperspectral embedding, token organization, and center--global prediction pathway form the task-specific adaptation studied here.

The main contributions are summarized as follows. (1) We establish a vision TTT-based spectral--spatial framework for HSI center-pixel classification, maintaining spatial correspondence throughout the token-processing stages. (2) We introduce a spectral-aware embedding pathway that exploits local spectral continuity while generating spatially aligned features suitable for ViT$^3$ mixing. (3) We evaluate the framework across four public HSI datasets and ten random seeds, and examine the roles of the TTT inner update, input embedding, center--global feature aggregation, spatial context size, and spatially separated sampling. The controlled inner-update comparisons provide a direct assessment of the patch-conditioned learning mechanism.

\section{Related work}
\label{sec:related}
\subsection{Spectral--spatial feature learning in hyperspectral classification}
HSI classification methods must combine rich spectral descriptions with spatial relationships. Earlier deep spectral--spatial architectures typically encode a neighborhood centered at the target pixel and assign the resulting feature vector to the center label. The spectral--spatial residual network (SSRN) uses consecutive spectral and spatial residual blocks in a 3-D convolutional architecture~\cite{zhong2018ssrn}. HybridSN subsequently couples 3-D convolutional feature extraction with 2-D convolution to learn joint representations at different stages~\cite{roy2020hybridsn}. These methods demonstrate the utility of treating local spectral structure and image-space context as complementary sources of information.

A local convolutional receptive field, however, is not the only means of representing dependencies across an HSI patch. Motivated by attention-based sequence modeling~\cite{vaswani2017attention}, SpectralFormer represents spectral signatures as grouped embeddings and incorporates spectral information flow across its Transformer layers~\cite{hong2022spectralformer}. SSFTT employs spectral--spatial feature tokenization before Transformer encoding, thereby constructing compact tokens from convolutional spatial--spectral features~\cite{sun2022ssftt}. More recently, the dual selective fusion Transformer network (DSFormer) has used selective fusion of spatial--spectral features and contextual tokens to capture information across different receptive fields~\cite{xu2025dsformer}. HyperTransXNet has explored local--global dynamics through hybrid spectral--spatial processing and dynamic token interactions~\cite{hypertransxnet2025}. The corresponding design space encompasses both the representation of spectral bands and the organization of spatial neighborhoods.

\subsection{State-space and alternative token mixers}
Attention-based methods have encouraged parallel work on sequence-processing operators with different computational trade-offs. The selective state-space model Mamba provides an input-dependent state transition mechanism for sequential representations~\cite{gu2024mamba}. In the HSI domain, S$^2$Mamba constructs complementary spatial and spectral processing pathways and combines their outputs through a spatial--spectral mixture gate~\cite{wang2025s2mamba}. Spatial-frequency joint learning Mamba (SFMamba) further explores spatial and frequency-domain information in hyperspectral classification~\cite{meng2026sfmamba}. These approaches, together with Transformer and hybrid convolutional neural network (CNN) architectures, constitute relevant baselines for evaluating another form of token interaction.

A central distinction of TTT-based mixing is that the current input sequence is used to construct temporary inner-model parameters rather than only to produce activations through a fixed mixing operator. Thus, TTT permits the model state used for query processing to depend on the content of the observed patch. In this work, the practical benefit of this mechanism is examined through comparisons with both a removed-mixer variant and a frozen-inner-model variant.

\subsection{Test-time training versus test-time adaptation}
In conventional test-time adaptation (TTA), a trained classifier can update selected model parameters using unlabeled target-domain observations, often to address corruption or distribution shift. HyperTTA, for example, studies hyperspectral classification under several forms of image degradation and adapts normalization parameters using confidence-aware entropy minimization~\cite{yue2026hypertta}. Its purpose is to improve performance when test inputs depart from the source-domain distribution.

Test-time training in the sense of Sun et al.~\cite{sun2025ttt} and ViT$^3$~\cite{han2026vit3} serves a different architectural role. An unlabeled inner objective is evaluated on key--value representations of the present sequence; the resulting fast weights process queries, while shared outer parameters remain unchanged during inference. In ViTTT-HSI, fast weights are reinitialized from their learned initial values for every patch. There is no persistent adaptation across patches, and no target class annotation participates in the inner update. The present study therefore focuses on TTT as an input-conditioned spatial token mixer for center-pixel classification rather than as a mechanism for adapting to external domain shifts.

\section{Methodology}
\label{sec:method}
\subsection{Problem formulation and framework overview}
\begin{figure}[!htbp]
\centering
\includegraphics[width=\textwidth]{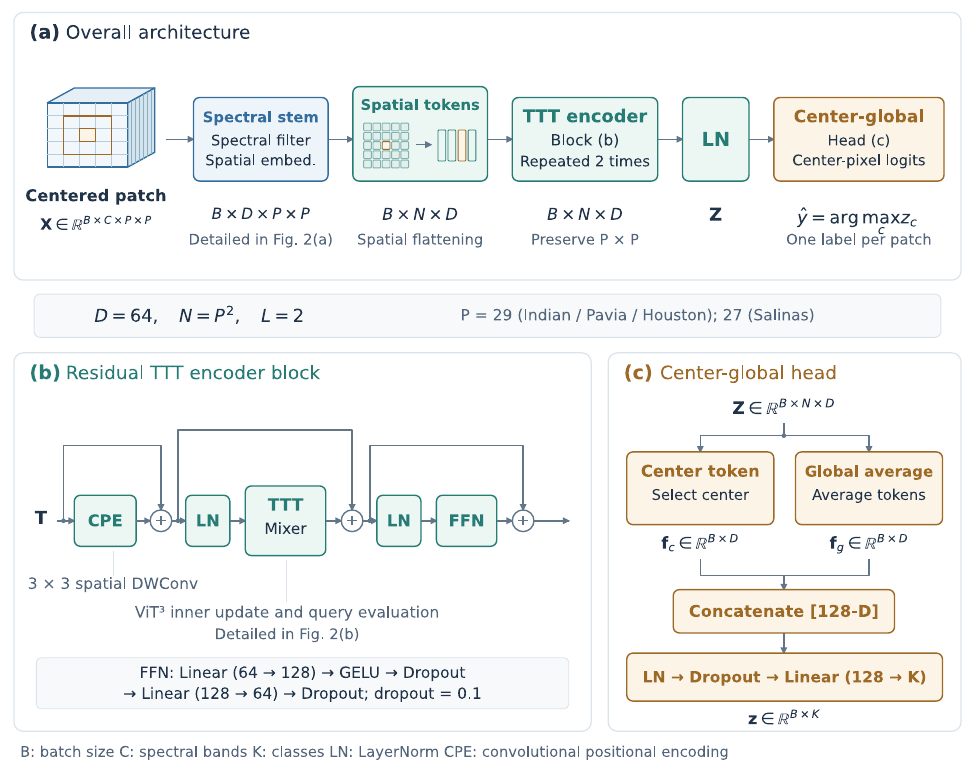}
\caption{Overall architecture of ViTTT-HSI. (a) A centered hyperspectral patch is encoded without spatial downsampling, processed by two TTT encoder blocks, and classified using a center--global feature head. (b) Each encoder block uses residual convolutional positional encoding (CPE), a pre-normalized TTT mixer inherited from ViT$^3$~\cite{han2026vit3}, and a pre-normalized feed-forward network. (c) The center token and spatial average are concatenated for center-pixel prediction. $C$, $D$, $N$, and $K$ denote spectral bands, token dimension, token count, and classes.}
\label{fig:framework}
\end{figure}

Let $\mathbf{I}\in\mathbb{R}^{H\times W\times C}$ represent an HSI cube, where $C$ is the number of spectral bands. For a labeled center coordinate $p$, an odd-sized neighborhood $\mathbf{X}_p\in\mathbb{R}^{C\times P\times P}$ is extracted. The goal is to predict the center label $y_p\in\{1,\dots,K\}$ without using ground-truth labels from other pixels in the patch during inner adaptation. Figure~\ref{fig:framework} summarizes the architecture. For a batch of $B$ patches, the embedding stage produces $\mathbf{T}^{(0)}\in\mathbb{R}^{B\times N\times D}$, where $N=P^2$ and $D=64$. The spatial tokens pass through $L=2$ encoder blocks, are normalized, and then feed a center--global classifier.

\subsection{Spectral-aware feature embedding}
\label{sec:stem}
Because the spectral axis has an intrinsic ordering, the embedding stem applies a convolution along spectral bands, without convolving over spatial coordinates in its first stage. A batch $\mathbf{X}\in\mathbb{R}^{B\times C\times P\times P}$ is reshaped to $\mathbb{R}^{B\times1\times C\times P\times P}$. An eight-channel 3-D convolution with kernel $(7,1,1)$, stride $(4,1,1)$, and spectral padding $3$ is followed by a Gaussian error linear unit (GELU) and adaptive average pooling to 16 spectral bins. This operation yields $\mathbb{R}^{B\times8\times16\times P\times P}$.

Folding the channel and pooled spectral axes produces 128 channels at each spatial position. A pointwise $1\times1$ convolution maps these features to $D=64$ channels, followed by batch normalization and GELU. A spatial depthwise $3\times3$ convolution and another GELU further refine local spatial context. The resulting $\mathbf{F}\in\mathbb{R}^{B\times D\times P\times P}$ is flattened only along its spatial axes:
\begin{equation}
\mathbf{T}^{(0)}=\operatorname{Flatten}_{HW}(\operatorname{Stem}(\mathbf{X}))^\top
\in\mathbb{R}^{B\times P^2\times D}.
\label{eq:stem}
\end{equation}
The superscript $\top$ in Eq.~\eqref{eq:stem} indicates the channel-to-feature-axis permutation, not a transposition of the full batch tensor. No spatial token is discarded and no class token is introduced. The linear-stem ablation replaces spectral convolution and pooling with a direct channel projection while retaining the remaining spatial embedding pathway.

\begin{figure}[!htbp]
\centering
\includegraphics[width=\textwidth]{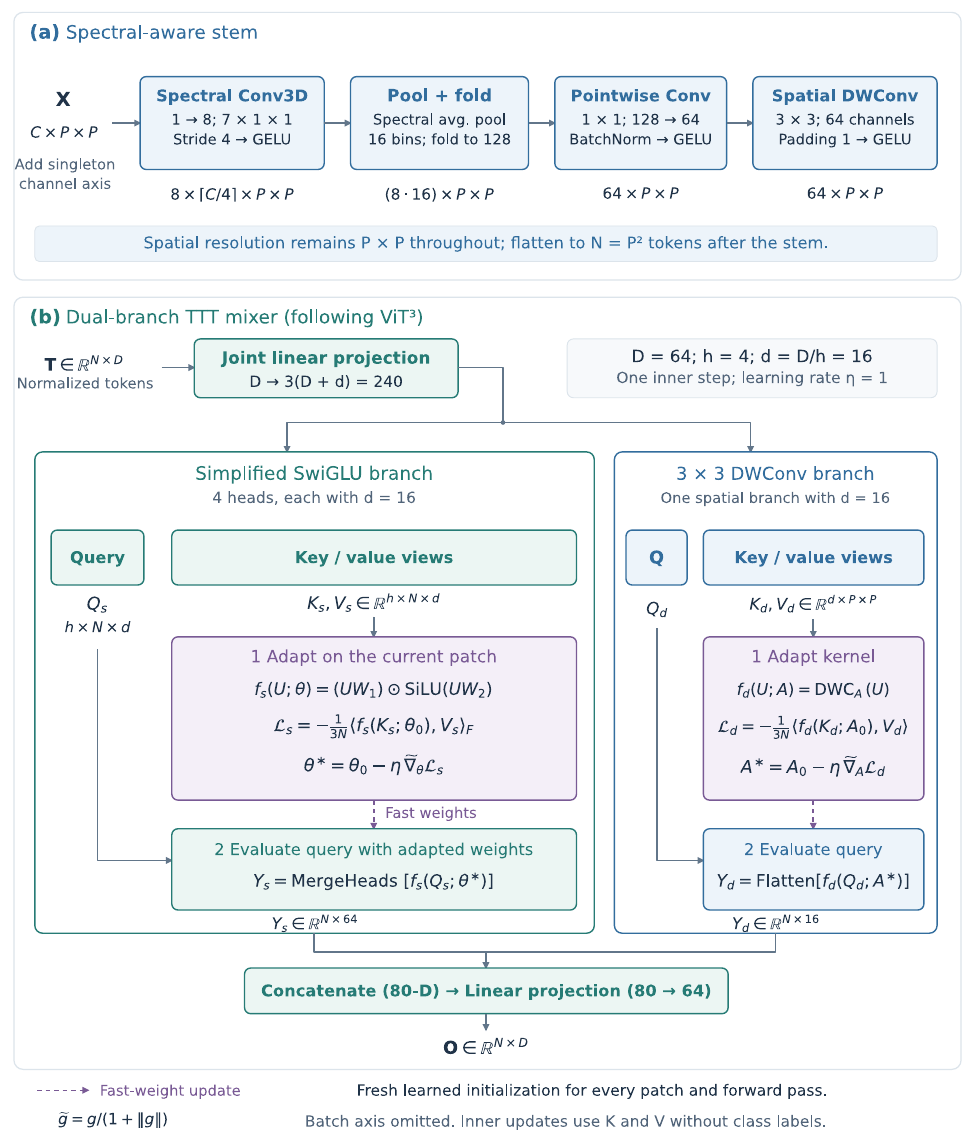}
\caption{Implementation details of the spectral-aware stem and ViT$^3$-based TTT mixer. (a) Spectral convolution, pooling, channel folding, and spatial depthwise convolution produce 64-dimensional spatial tokens. (b) Four 16-dimensional simplified swish-gated linear unit (SwiGLU) heads and a separate 16-channel depthwise convolution branch each perform a one-step, unlabeled key--value inner update and then evaluate query features using the patch-specific fast weights. Both branch designs and their analytical update mechanism follow ViT$^3$~\cite{han2026vit3}. Fast weights do not persist between patches.}
\label{fig:stem-ttt}
\end{figure}

\subsection{TTT-based spatial encoder}
\label{sec:ttt}
Each encoder block first introduces a residual convolutional positional encoding, $\widehat{\mathbf{T}}=\mathbf{T}+\operatorname{Flatten}_{HW}(\mathrm{DWConv}_{3\times3}(\mathbf{T}))$. The subsequent pre-normalized token mixer and feed-forward sublayer are residual:
\begin{align}
\mathbf{U} &= \widehat{\mathbf{T}}+\operatorname{TTT}(\operatorname{LN}(\widehat{\mathbf{T}})),\\
\mathbf{T}' &= \mathbf{U}+\operatorname{FFN}(\operatorname{LN}(\mathbf{U})).
\label{eq:block}
\end{align}
Here, DWConv denotes depthwise convolution, LN denotes layer normalization, and FFN denotes a feed-forward network. The feed-forward network uses two linear transformations with dimensions $64\rightarrow128\rightarrow64$, GELU, and dropout with probability $0.1$.

The TTT mixer is the original ViT$^3$ dual-branch construction~\cite{han2026vit3}. A joint projection produces distinct query (Q), key (K), and value (V) views for a simplified SwiGLU branch and a depthwise convolution branch. Four gated heads each have dimension $d=D/4=16$, while the convolution branch has 16 channels. Their projected output dimensions are 64 and 16, respectively, and a final linear projection maps the 80 concatenated features to 64 dimensions.

For each gated head, denote the learnable inner initialization by $\Theta_0=\{\mathbf{W}_1,\mathbf{W}_2\}$. The inner network is
\begin{equation}
f_{\Theta}(\mathbf{z})=(\mathbf{z}\mathbf{W}_1)\odot\operatorname{SiLU}(\mathbf{z}\mathbf{W}_2).
\label{eq:swiglu}
\end{equation}
Here, SiLU denotes the sigmoid linear unit. For $N$ spatial tokens, a compact expression for the negative dot-product inner objective is
\begin{equation}
\mathcal{L}_{\mathrm{inner}}(\Theta;\mathbf{K},\mathbf{V})=
-\frac{s}{N}\sum_{i=1}^{N}\langle f_{\Theta}(\mathbf{K}_i),\mathbf{V}_i\rangle,
\qquad s=1/3.
\label{eq:inner}
\end{equation}
The convolution inner model uses the corresponding spatial dot-product objective with a $3\times3$ depthwise kernel. In the implementation, gradients are computed analytically without explicitly materializing the scalar objective, and each parameter gradient is rescaled by its norm along the implementation-defined axes:
\begin{equation}
\widetilde{\mathbf{G}}=\frac{\mathbf{G}}{1+\|\mathbf{G}\|},\quad
\Theta_p^*=\Theta_0-\eta\widetilde{\mathbf{G}}_p,
\quad \eta=1.
\label{eq:innerupdate}
\end{equation}
Equation~\eqref{eq:innerupdate} represents the one-step update for the current patch $p$. The gated and convolutional gradients are computed separately and normalized over their respective tensor axes. This operation is repeated independently for each inner branch and each input patch. The resulting temporary weights produce query outputs, $\mathbf{O}_p=f_{\Theta_p^*}(\mathbf{Q}_p)$; the outer-network initializations remain shared across patches and are not overwritten at inference. Crucially, this inner update uses no center-pixel labels or test-set annotations.

\subsection{Center--global head and outer optimization}
After two encoder blocks, a final layer normalization produces $\mathbf{Z}\in\mathbb{R}^{B\times N\times D}$. Let $i_c=\lfloor P/2\rfloor P+\lfloor P/2\rfloor$ be the center-token index under zero-based indexing. We define
\begin{align}
\mathbf{f}_c&=\mathbf{Z}_{:,i_c,:},&
\mathbf{f}_g&=\frac{1}{N}\sum_{i=1}^{N}\mathbf{Z}_{:,i,:},\\
\widehat{\mathbf{y}}&=\mathrm{Linear}(\mathrm{Dropout}(\mathrm{LN}([\mathbf{f}_c;\mathbf{f}_g]))).
\label{eq:head}
\end{align}
The global average includes the center token. The classifier receives 128 features and produces $K$ class logits. Outer-network parameters, including the initial inner weights, are optimized end-to-end using standard supervised cross-entropy on training center-pixel labels.

\FloatBarrier
\section{Experimental setup}
\label{sec:setup}
\subsection{Datasets and sampling protocol}
Four widely used HSI benchmarks are considered: Indian Pines (16 classes), Pavia University (9 classes), Salinas (16 classes), and Houston 2013 (15 classes). For each class, a seed-controlled random partition of labeled center pixels is used. Indian Pines and Houston 2013 each allocate 5\% of labeled pixels to training and another 5\% to validation; the rest is reserved for testing. Pavia University and Salinas each allocate 1\% to training and 1\% to validation. Dataset-specific rounding and a minimum of one sample per class are implemented where applicable. The ten seeds are 3405--3414. These experiments use random pixel splits rather than the official Houston 2013 fixed training/test division. Training-pixel spectra alone determine the per-band normalization statistics.

The input patch sizes are 29 for Indian Pines, Pavia University, and Houston 2013, and 27 for Salinas. All sampling protocols are applied to labeled center pixels, with disjoint training, validation, and test center-pixel sets. A separate spatial-block experiment further examines classification with spatially separated samples.

\subsection{Implementation and checkpoint selection}
The model has embedding dimension 64, encoder depth 2, four gated TTT heads and one convolution TTT branch, a feed-forward expansion factor of 2, 16 pooled spectral bins, and dropout 0.1. Training uses PyTorch 2.2.0, a single NVIDIA RTX 3090, AdamW with initial learning rate $3\times10^{-4}$ and weight decay 0.01, cosine annealing, batch size 32, and gradient-norm clipping at 1.0. A maximum of 300 epochs is allowed, with early-stopping patience 50.

The primary checkpoint criterion is the highest validation overall accuracy (OA).\label{sec:checkpoint} When multiple epochs attain the same maximum number of correct validation predictions, the model with the lower validation cross-entropy is retained. Early-stopping patience resets only when validation OA strictly increases. Test labels are not used for checkpoint selection or hyperparameter updates. Reported results are the mean and sample standard deviation across ten seeds, unless stated otherwise.

\subsection{Evaluation criteria and baselines}
We report overall accuracy (OA), average accuracy (AA, the mean per-class recall), Cohen's kappa, macro-F1, weighted-F1, and per-class recall. The comparison methods are SFMamba~\cite{meng2026sfmamba}, HyperTransXNet~\cite{hypertransxnet2025}, S$^2$Mamba~\cite{wang2025s2mamba}, DSFormer~\cite{xu2025dsformer}, and SSFTT~\cite{sun2022ssftt}. All methods follow the same class-stratified random-sampling protocol, using seeds 3405--3414 and the training and validation fractions specified in Sec.~\ref{sec:setup}. Each competing method uses the best-performing hyperparameter configuration specified in its original paper or official open-source implementation. Results are summarized over ten runs as the mean and standard deviation.

\FloatBarrier
\section{Results and discussion}
\label{sec:results}
\subsection{Overall and class-wise classification performance}
Tables~\ref{tab:comparison-indian}--\ref{tab:comparison-houston13} present class-wise recall, overall accuracy (OA), average accuracy (AA), Cohen's kappa, macro-F1, and weighted-F1 for the five comparison methods and ViTTT-HSI. Each entry reports the mean and standard deviation over ten random seeds. The mean OA values of ViTTT-HSI are 97.6390\% on Indian Pines, 98.8753\% on Pavia University, 99.3292\% on Salinas, and 98.4283\% on Houston 2013.

On Indian Pines, ViTTT-HSI exceeds the strongest comparison method, DSFormer, by 1.8470 percentage points in mean OA. The corresponding improvements over the highest competing mean OA are 0.3315 percentage points on Pavia University, 0.3636 on Salinas, and 0.2728 on Houston 2013. The class-wise results further characterize performance across land-cover categories. For example, on Indian Pines, ViTTT-HSI achieves recalls of 85.38\% and 72.78\% for classes 7 and 9, respectively, while other methods show lower mean recall on these challenging classes. On Pavia University, the method achieves 99.23\% recall for class 7 and 99.56\% for class 9. Salinas and Houston 2013 also show strong recall across most classes. The detailed per-class comparison allows the OA and AA differences to be interpreted alongside class-specific performance.

\begin{table}[!htbp]
\caption{Class-wise recall and overall performance on Indian Pines (\%, ten-seed mean $\pm$ standard deviation). All metrics, including kappa, are expressed as percentages. Class IDs follow the reference labels; bold marks the highest mean in each row.}
\label{tab:comparison-indian}
\centering
\begingroup\singlespacing\renewcommand{\arraystretch}{0.85}\setlength{\tabcolsep}{2.8pt}
\resizebox{\textwidth}{!}{%
\begin{tabular}{lrrrrrr}
\toprule
Metric & SFMamba & HyperTransXNet & S$^2$Mamba & DSFormer & SSFTT & ViTTT-HSI \\
\midrule
Class 01 & 71.19 $\pm$ 25.21 & 67.86 $\pm$ 20.05 & 0.48 $\pm$ 1.00 & 79.05 $\pm$ 18.87 & 59.05 $\pm$ 28.14 & \textbf{96.43 $\pm$ 2.31} \\
Class 02 & 91.73 $\pm$ 3.88 & 93.68 $\pm$ 2.91 & 93.17 $\pm$ 2.57 & 92.65 $\pm$ 1.91 & 91.76 $\pm$ 2.50 & \textbf{96.01 $\pm$ 2.21} \\
Class 03 & 92.44 $\pm$ 3.46 & 93.14 $\pm$ 2.97 & 93.14 $\pm$ 3.55 & \textbf{96.88 $\pm$ 2.33} & 94.50 $\pm$ 2.23 & 96.74 $\pm$ 1.48 \\
Class 04 & 93.00 $\pm$ 5.47 & 95.16 $\pm$ 4.01 & 91.74 $\pm$ 6.72 & 93.80 $\pm$ 4.07 & 93.94 $\pm$ 7.06 & \textbf{98.22 $\pm$ 1.71} \\
Class 05 & 92.48 $\pm$ 4.78 & 93.52 $\pm$ 3.04 & 91.68 $\pm$ 3.52 & 93.82 $\pm$ 2.58 & 93.36 $\pm$ 3.01 & \textbf{93.93 $\pm$ 3.32} \\
Class 06 & 98.34 $\pm$ 1.15 & 97.07 $\pm$ 2.22 & 99.59 $\pm$ 0.28 & 98.86 $\pm$ 1.05 & 98.63 $\pm$ 0.96 & \textbf{99.60 $\pm$ 0.28} \\
Class 07 & 48.85 $\pm$ 35.44 & 30.00 $\pm$ 25.37 & 0.77 $\pm$ 1.62 & 73.08 $\pm$ 18.13 & 32.69 $\pm$ 31.47 & \textbf{85.38 $\pm$ 15.89} \\
Class 08 & \textbf{100.00 $\pm$ 0.00} & 99.93 $\pm$ 0.16 & \textbf{100.00 $\pm$ 0.00} & 99.81 $\pm$ 0.59 & 98.14 $\pm$ 1.65 & \textbf{100.00 $\pm$ 0.00} \\
Class 09 & 42.78 $\pm$ 18.71 & 49.44 $\pm$ 23.49 & 0.00 $\pm$ 0.00 & 54.44 $\pm$ 19.39 & 31.67 $\pm$ 24.85 & \textbf{72.78 $\pm$ 25.32} \\
Class 10 & 89.67 $\pm$ 3.51 & 89.79 $\pm$ 2.99 & 91.67 $\pm$ 3.62 & 90.50 $\pm$ 3.42 & 91.51 $\pm$ 2.72 & \textbf{94.77 $\pm$ 2.70} \\
Class 11 & 96.23 $\pm$ 1.73 & 96.89 $\pm$ 1.90 & 97.22 $\pm$ 1.59 & 97.49 $\pm$ 1.48 & 95.96 $\pm$ 2.28 & \textbf{98.89 $\pm$ 1.12} \\
Class 12 & 88.14 $\pm$ 3.44 & 90.00 $\pm$ 4.01 & 95.98 $\pm$ 2.60 & 93.81 $\pm$ 2.13 & 85.40 $\pm$ 7.02 & \textbf{96.04 $\pm$ 1.81} \\
Class 13 & 97.89 $\pm$ 2.55 & 95.35 $\pm$ 6.01 & 99.46 $\pm$ 0.92 & 98.97 $\pm$ 0.86 & 99.08 $\pm$ 1.51 & \textbf{99.78 $\pm$ 0.38} \\
Class 14 & 98.01 $\pm$ 1.41 & 96.98 $\pm$ 2.14 & 98.13 $\pm$ 1.25 & 98.67 $\pm$ 0.84 & 97.88 $\pm$ 1.74 & \textbf{99.95 $\pm$ 0.09} \\
Class 15 & 94.17 $\pm$ 3.35 & 90.23 $\pm$ 4.40 & 89.08 $\pm$ 6.39 & 98.51 $\pm$ 1.38 & 93.99 $\pm$ 4.47 & \textbf{99.20 $\pm$ 2.06} \\
Class 16 & \textbf{95.30 $\pm$ 6.36} & 83.73 $\pm$ 9.82 & 92.17 $\pm$ 7.32 & 95.06 $\pm$ 3.91 & 89.28 $\pm$ 10.93 & 95.18 $\pm$ 5.90 \\
\midrule
OA & 94.10 $\pm$ 1.06 & 94.22 $\pm$ 0.81 & 94.54 $\pm$ 0.65 & 95.79 $\pm$ 0.64 & 94.04 $\pm$ 0.90 & \textbf{97.64 $\pm$ 0.58} \\
AA & 86.89 $\pm$ 3.01 & 85.17 $\pm$ 2.11 & 77.14 $\pm$ 0.98 & 90.96 $\pm$ 1.36 & 84.18 $\pm$ 3.68 & \textbf{95.18 $\pm$ 1.83} \\
Kappa & 93.27 $\pm$ 1.21 & 93.41 $\pm$ 0.92 & 93.77 $\pm$ 0.74 & 95.20 $\pm$ 0.73 & 93.20 $\pm$ 1.02 & \textbf{97.31 $\pm$ 0.66} \\
Macro-F1 & 87.90 $\pm$ 2.92 & 86.85 $\pm$ 2.62 & 77.07 $\pm$ 0.81 & 92.47 $\pm$ 1.48 & 84.74 $\pm$ 3.64 & \textbf{95.47 $\pm$ 1.69} \\
Weighted-F1 & 94.01 $\pm$ 1.08 & 94.13 $\pm$ 0.80 & 94.09 $\pm$ 0.66 & 95.75 $\pm$ 0.64 & 93.91 $\pm$ 0.91 & \textbf{97.63 $\pm$ 0.59} \\
\bottomrule
\end{tabular}%
}
\endgroup
\par\vspace{12pt}
\caption{Class-wise recall and overall performance on Pavia University (\%, ten-seed mean $\pm$ standard deviation). All metrics, including kappa, are expressed as percentages. Class IDs follow the reference labels; bold marks the highest mean in each row.}
\label{tab:comparison-pavia}
\centering
\begingroup\singlespacing\renewcommand{\arraystretch}{0.85}\setlength{\tabcolsep}{2.8pt}
\resizebox{\textwidth}{!}{%
\begin{tabular}{lrrrrrr}
\toprule
Metric & SFMamba & HyperTransXNet & S$^2$Mamba & DSFormer & SSFTT & ViTTT-HSI \\
\midrule
Class 01 & 97.05 $\pm$ 1.44 & 97.41 $\pm$ 1.53 & \textbf{98.67 $\pm$ 0.94} & 97.79 $\pm$ 1.22 & 97.09 $\pm$ 1.46 & 98.30 $\pm$ 0.94 \\
Class 02 & 99.78 $\pm$ 0.14 & 99.50 $\pm$ 0.43 & \textbf{99.82 $\pm$ 0.16} & 99.82 $\pm$ 0.11 & 99.79 $\pm$ 0.20 & 99.73 $\pm$ 0.29 \\
Class 03 & 94.16 $\pm$ 1.65 & 91.44 $\pm$ 5.70 & 94.21 $\pm$ 3.46 & 94.53 $\pm$ 2.76 & 90.11 $\pm$ 4.10 & \textbf{95.11 $\pm$ 5.75} \\
Class 04 & 95.59 $\pm$ 1.52 & 92.32 $\pm$ 3.10 & 96.86 $\pm$ 1.07 & 96.56 $\pm$ 1.21 & 94.90 $\pm$ 2.78 & \textbf{98.06 $\pm$ 0.95} \\
Class 05 & 99.79 $\pm$ 0.18 & 99.24 $\pm$ 0.80 & 99.98 $\pm$ 0.05 & \textbf{99.99 $\pm$ 0.02} & 99.85 $\pm$ 0.31 & 99.96 $\pm$ 0.10 \\
Class 06 & \textbf{99.31 $\pm$ 0.69} & 98.31 $\pm$ 0.92 & 98.52 $\pm$ 1.14 & 99.01 $\pm$ 0.99 & 98.75 $\pm$ 1.72 & 98.86 $\pm$ 2.37 \\
Class 07 & 94.06 $\pm$ 5.39 & 93.03 $\pm$ 3.86 & 93.44 $\pm$ 7.29 & 95.79 $\pm$ 6.94 & 99.01 $\pm$ 0.73 & \textbf{99.23 $\pm$ 1.87} \\
Class 08 & 96.30 $\pm$ 1.83 & 96.50 $\pm$ 2.35 & \textbf{98.48 $\pm$ 1.64} & 95.35 $\pm$ 2.58 & 89.88 $\pm$ 4.86 & 97.75 $\pm$ 1.37 \\
Class 09 & 91.86 $\pm$ 6.17 & 92.69 $\pm$ 3.89 & 93.08 $\pm$ 5.32 & 95.46 $\pm$ 5.91 & 92.17 $\pm$ 6.54 & \textbf{99.56 $\pm$ 0.66} \\
\midrule
OA & 98.07 $\pm$ 0.38 & 97.51 $\pm$ 0.51 & 98.54 $\pm$ 0.39 & 98.31 $\pm$ 0.54 & 97.38 $\pm$ 0.53 & \textbf{98.88 $\pm$ 0.49} \\
AA & 96.43 $\pm$ 1.09 & 95.61 $\pm$ 0.96 & 97.01 $\pm$ 1.20 & 97.14 $\pm$ 1.38 & 95.73 $\pm$ 1.00 & \textbf{98.51 $\pm$ 0.72} \\
Kappa & 97.45 $\pm$ 0.50 & 96.69 $\pm$ 0.68 & 98.07 $\pm$ 0.52 & 97.76 $\pm$ 0.72 & 96.53 $\pm$ 0.70 & \textbf{98.51 $\pm$ 0.65} \\
Macro-F1 & 96.65 $\pm$ 0.84 & 95.92 $\pm$ 0.91 & 97.53 $\pm$ 0.98 & 97.35 $\pm$ 1.16 & 95.94 $\pm$ 1.01 & \textbf{98.47 $\pm$ 0.76} \\
Weighted-F1 & 98.07 $\pm$ 0.38 & 97.49 $\pm$ 0.52 & 98.53 $\pm$ 0.40 & 98.31 $\pm$ 0.55 & 97.36 $\pm$ 0.54 & \textbf{98.87 $\pm$ 0.50} \\
\bottomrule
\end{tabular}%
}
\endgroup
\end{table}

\begin{table}[!htbp]
\caption{Class-wise recall and overall performance on Salinas (\%, ten-seed mean $\pm$ standard deviation). All metrics, including kappa, are expressed as percentages. Class IDs follow the reference labels; bold marks the highest mean in each row.}
\label{tab:comparison-salinas}
\centering
\begingroup\singlespacing\renewcommand{\arraystretch}{0.85}\setlength{\tabcolsep}{2.8pt}
\resizebox{\textwidth}{!}{%
\begin{tabular}{lrrrrrr}
\toprule
Metric & SFMamba & HyperTransXNet & S$^2$Mamba & DSFormer & SSFTT & ViTTT-HSI \\
\midrule
Class 01 & 98.70 $\pm$ 2.79 & 99.36 $\pm$ 0.82 & \textbf{99.90 $\pm$ 0.24} & 97.64 $\pm$ 5.07 & 99.84 $\pm$ 0.30 & 99.85 $\pm$ 0.41 \\
Class 02 & 99.99 $\pm$ 0.03 & 99.96 $\pm$ 0.11 & \textbf{100.00 $\pm$ 0.00} & \textbf{100.00 $\pm$ 0.00} & 99.91 $\pm$ 0.16 & 99.99 $\pm$ 0.02 \\
Class 03 & 98.69 $\pm$ 3.80 & 99.16 $\pm$ 1.15 & \textbf{99.98 $\pm$ 0.04} & 99.78 $\pm$ 0.53 & \textbf{99.98 $\pm$ 0.07} & 99.53 $\pm$ 0.50 \\
Class 04 & 98.95 $\pm$ 2.13 & 97.56 $\pm$ 2.60 & 98.66 $\pm$ 1.24 & \textbf{99.43 $\pm$ 0.33} & 99.02 $\pm$ 0.70 & 99.39 $\pm$ 1.34 \\
Class 05 & 98.69 $\pm$ 1.92 & 98.55 $\pm$ 1.02 & 98.64 $\pm$ 2.20 & 98.91 $\pm$ 1.60 & \textbf{99.02 $\pm$ 1.23} & 98.01 $\pm$ 0.83 \\
Class 06 & 99.89 $\pm$ 0.20 & 99.83 $\pm$ 0.36 & 99.83 $\pm$ 0.53 & \textbf{100.00 $\pm$ 0.00} & 99.89 $\pm$ 0.19 & 99.90 $\pm$ 0.23 \\
Class 07 & 99.92 $\pm$ 0.09 & 99.64 $\pm$ 0.71 & 99.97 $\pm$ 0.05 & \textbf{99.99 $\pm$ 0.01} & 99.78 $\pm$ 0.58 & 99.86 $\pm$ 0.29 \\
Class 08 & 92.57 $\pm$ 2.78 & 95.91 $\pm$ 2.55 & 96.23 $\pm$ 1.44 & 96.88 $\pm$ 0.72 & 98.36 $\pm$ 0.83 & \textbf{99.12 $\pm$ 0.64} \\
Class 09 & 99.98 $\pm$ 0.03 & 99.79 $\pm$ 0.30 & \textbf{100.00 $\pm$ 0.00} & \textbf{100.00 $\pm$ 0.00} & 99.97 $\pm$ 0.04 & 99.98 $\pm$ 0.06 \\
Class 10 & 97.93 $\pm$ 1.52 & 99.08 $\pm$ 0.77 & 98.95 $\pm$ 1.18 & 98.32 $\pm$ 1.14 & 98.76 $\pm$ 1.25 & \textbf{99.44 $\pm$ 0.87} \\
Class 11 & 99.01 $\pm$ 0.74 & 99.83 $\pm$ 0.25 & 99.52 $\pm$ 0.09 & \textbf{99.94 $\pm$ 0.15} & 99.66 $\pm$ 0.60 & 99.34 $\pm$ 1.09 \\
Class 12 & 99.25 $\pm$ 1.67 & 99.38 $\pm$ 1.11 & \textbf{99.97 $\pm$ 0.06} & 99.88 $\pm$ 0.12 & 99.76 $\pm$ 0.44 & 99.86 $\pm$ 0.19 \\
Class 13 & 99.65 $\pm$ 0.46 & 99.43 $\pm$ 1.17 & 99.62 $\pm$ 1.12 & \textbf{99.97 $\pm$ 0.08} & 96.34 $\pm$ 7.32 & 98.15 $\pm$ 2.77 \\
Class 14 & 98.91 $\pm$ 0.97 & 98.48 $\pm$ 1.72 & 98.95 $\pm$ 1.02 & \textbf{99.39 $\pm$ 0.60} & 97.76 $\pm$ 2.80 & 97.67 $\pm$ 2.08 \\
Class 15 & 92.19 $\pm$ 2.19 & 94.50 $\pm$ 2.69 & 94.06 $\pm$ 3.33 & 95.36 $\pm$ 2.18 & 97.53 $\pm$ 1.60 & \textbf{98.93 $\pm$ 0.73} \\
Class 16 & 98.84 $\pm$ 1.97 & \textbf{99.58 $\pm$ 0.75} & 99.49 $\pm$ 0.52 & 99.04 $\pm$ 2.49 & 98.48 $\pm$ 0.88 & 98.29 $\pm$ 3.55 \\
\midrule
OA & 96.96 $\pm$ 0.40 & 98.02 $\pm$ 0.60 & 98.18 $\pm$ 0.69 & 98.41 $\pm$ 0.42 & 98.97 $\pm$ 0.28 & \textbf{99.33 $\pm$ 0.27} \\
AA & 98.32 $\pm$ 0.42 & 98.75 $\pm$ 0.34 & 98.99 $\pm$ 0.41 & 99.03 $\pm$ 0.41 & 99.00 $\pm$ 0.57 & \textbf{99.21 $\pm$ 0.44} \\
Kappa & 96.62 $\pm$ 0.44 & 97.80 $\pm$ 0.67 & 97.97 $\pm$ 0.77 & 98.23 $\pm$ 0.46 & 98.85 $\pm$ 0.32 & \textbf{99.25 $\pm$ 0.30} \\
Macro-F1 & 98.23 $\pm$ 0.46 & 98.67 $\pm$ 0.39 & 98.98 $\pm$ 0.40 & 98.97 $\pm$ 0.40 & 99.00 $\pm$ 0.51 & \textbf{99.19 $\pm$ 0.45} \\
Weighted-F1 & 96.96 $\pm$ 0.39 & 98.02 $\pm$ 0.60 & 98.18 $\pm$ 0.69 & 98.41 $\pm$ 0.42 & 98.97 $\pm$ 0.28 & \textbf{99.33 $\pm$ 0.27} \\
\bottomrule
\end{tabular}%
}
\endgroup
\par\vspace{12pt}
\caption{Class-wise recall and overall performance on Houston 2013 (\%, ten-seed mean $\pm$ standard deviation). All metrics, including kappa, are expressed as percentages. Class IDs follow the reference labels; bold marks the highest mean in each row.}
\label{tab:comparison-houston13}
\centering
\begingroup\singlespacing\renewcommand{\arraystretch}{0.85}\setlength{\tabcolsep}{2.8pt}
\resizebox{\textwidth}{!}{%
\begin{tabular}{lrrrrrr}
\toprule
Metric & SFMamba & HyperTransXNet & S$^2$Mamba & DSFormer & SSFTT & ViTTT-HSI \\
\midrule
Class 01 & 97.96 $\pm$ 1.08 & 97.63 $\pm$ 1.60 & 97.24 $\pm$ 1.02 & 98.18 $\pm$ 1.29 & 98.28 $\pm$ 1.63 & \textbf{98.41 $\pm$ 1.50} \\
Class 02 & 98.97 $\pm$ 0.97 & 98.32 $\pm$ 1.55 & 98.71 $\pm$ 0.53 & 99.23 $\pm$ 0.47 & \textbf{99.47 $\pm$ 0.76} & 99.20 $\pm$ 0.49 \\
Class 03 & 99.25 $\pm$ 0.43 & 99.36 $\pm$ 0.44 & 99.65 $\pm$ 0.30 & 98.82 $\pm$ 1.27 & 99.57 $\pm$ 0.29 & \textbf{100.00 $\pm$ 0.00} \\
Class 04 & 98.84 $\pm$ 1.08 & 97.07 $\pm$ 1.97 & 98.44 $\pm$ 0.65 & \textbf{99.23 $\pm$ 0.55} & 98.64 $\pm$ 0.50 & 99.09 $\pm$ 1.77 \\
Class 05 & 99.96 $\pm$ 0.09 & 99.96 $\pm$ 0.10 & 99.99 $\pm$ 0.03 & \textbf{100.00 $\pm$ 0.00} & \textbf{100.00 $\pm$ 0.00} & \textbf{100.00 $\pm$ 0.00} \\
Class 06 & 90.34 $\pm$ 5.67 & 93.17 $\pm$ 4.66 & 83.55 $\pm$ 4.00 & 93.89 $\pm$ 5.43 & 94.57 $\pm$ 5.95 & \textbf{97.34 $\pm$ 2.47} \\
Class 07 & 98.11 $\pm$ 1.09 & 95.95 $\pm$ 2.03 & \textbf{98.98 $\pm$ 0.71} & 98.48 $\pm$ 1.29 & 97.43 $\pm$ 2.79 & 97.94 $\pm$ 0.84 \\
Class 08 & 93.36 $\pm$ 2.28 & 92.83 $\pm$ 2.06 & 84.38 $\pm$ 4.50 & 94.50 $\pm$ 1.64 & 93.92 $\pm$ 3.86 & \textbf{94.64 $\pm$ 3.18} \\
Class 09 & 95.36 $\pm$ 2.23 & 94.80 $\pm$ 2.36 & 91.27 $\pm$ 2.54 & 95.36 $\pm$ 1.92 & 96.75 $\pm$ 1.36 & \textbf{97.31 $\pm$ 2.08} \\
Class 10 & 99.23 $\pm$ 0.75 & 99.10 $\pm$ 1.55 & 96.62 $\pm$ 1.44 & 99.25 $\pm$ 0.54 & 99.32 $\pm$ 0.75 & \textbf{99.84 $\pm$ 0.23} \\
Class 11 & 98.61 $\pm$ 1.16 & 97.76 $\pm$ 1.76 & \textbf{98.91 $\pm$ 1.24} & 98.88 $\pm$ 1.21 & 98.28 $\pm$ 1.71 & 98.74 $\pm$ 1.30 \\
Class 12 & 97.08 $\pm$ 1.24 & 95.98 $\pm$ 1.16 & 93.10 $\pm$ 3.55 & 97.12 $\pm$ 1.16 & \textbf{98.70 $\pm$ 0.65} & 97.97 $\pm$ 2.12 \\
Class 13 & 95.89 $\pm$ 3.44 & 92.53 $\pm$ 3.12 & 44.80 $\pm$ 32.17 & \textbf{97.40 $\pm$ 2.53} & 96.43 $\pm$ 3.32 & 96.36 $\pm$ 2.22 \\
Class 14 & \textbf{100.00 $\pm$ 0.00} & 99.30 $\pm$ 1.04 & \textbf{100.00 $\pm$ 0.00} & 99.61 $\pm$ 0.48 & \textbf{100.00 $\pm$ 0.00} & \textbf{100.00 $\pm$ 0.00} \\
Class 15 & 99.97 $\pm$ 0.11 & 99.85 $\pm$ 0.23 & \textbf{100.00 $\pm$ 0.00} & 99.95 $\pm$ 0.11 & 99.98 $\pm$ 0.05 & 99.97 $\pm$ 0.11 \\
\midrule
OA & 97.76 $\pm$ 0.27 & 97.02 $\pm$ 0.62 & 94.40 $\pm$ 1.28 & 98.08 $\pm$ 0.26 & 98.16 $\pm$ 0.59 & \textbf{98.43 $\pm$ 0.60} \\
AA & 97.53 $\pm$ 0.41 & 96.91 $\pm$ 0.66 & 92.38 $\pm$ 2.42 & 97.99 $\pm$ 0.50 & 98.09 $\pm$ 0.74 & \textbf{98.45 $\pm$ 0.58} \\
Kappa & 97.58 $\pm$ 0.29 & 96.78 $\pm$ 0.67 & 93.94 $\pm$ 1.39 & 97.92 $\pm$ 0.29 & 98.01 $\pm$ 0.64 & \textbf{98.30 $\pm$ 0.64} \\
Macro-F1 & 97.71 $\pm$ 0.27 & 96.96 $\pm$ 0.63 & 92.68 $\pm$ 2.88 & 98.11 $\pm$ 0.35 & 98.14 $\pm$ 0.67 & \textbf{98.44 $\pm$ 0.60} \\
Weighted-F1 & 97.75 $\pm$ 0.27 & 97.02 $\pm$ 0.62 & 93.95 $\pm$ 1.74 & 98.08 $\pm$ 0.27 & 98.15 $\pm$ 0.59 & \textbf{98.43 $\pm$ 0.60} \\
\bottomrule
\end{tabular}%
}
\endgroup
\end{table}

\clearpage

\subsection{Classification maps and error visualization}
\begin{figure}[!htbp]
\centering
\includegraphics[width=0.90\textwidth]{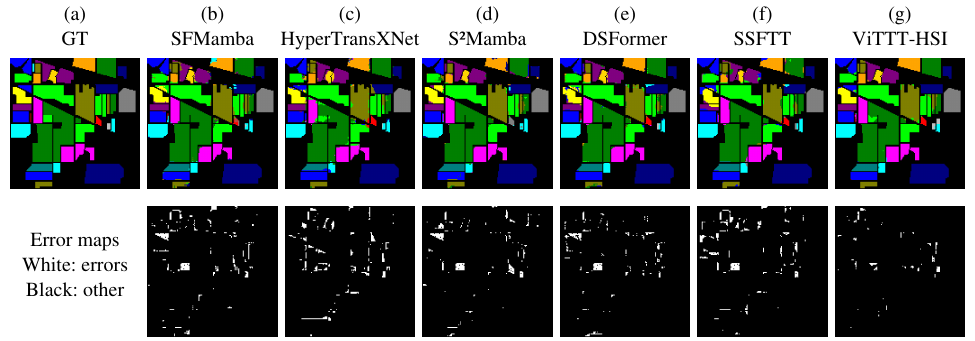}
\caption{Classification maps on Indian Pines: (a) ground truth; (b) SFMamba; (c) HyperTransXNet; (d) S$^2$Mamba; (e) DSFormer; (f) SSFTT; (g) ViTTT-HSI. Each model's prediction appears above its error map.}
\label{fig:map-indian}
\par\vspace{10pt}
\includegraphics[width=0.90\textwidth]{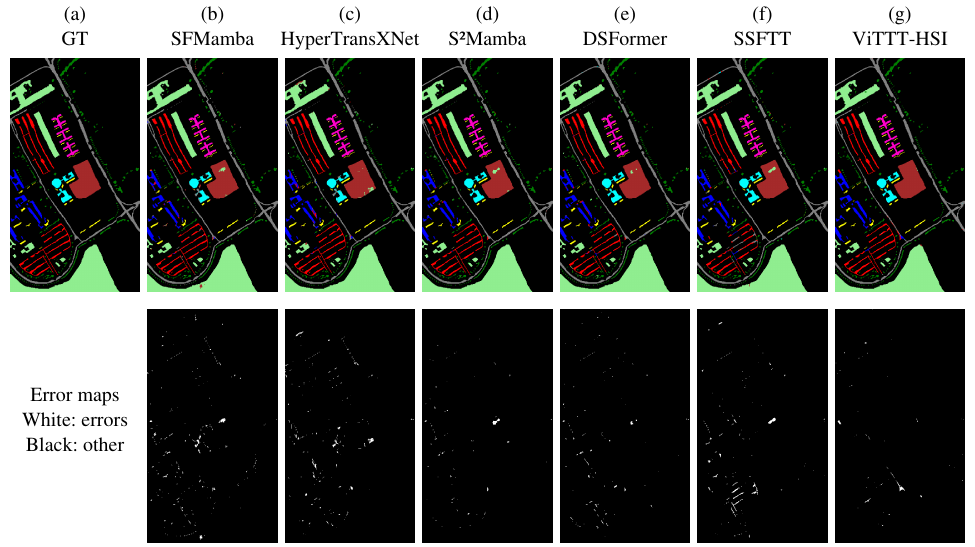}
\caption{Classification maps on Pavia University. Panel order and the error-map convention follow Fig.~\ref{fig:map-indian}.}
\label{fig:map-pavia}
\end{figure}

\begin{figure}[!htbp]
\centering
\includegraphics[width=0.84\textwidth]{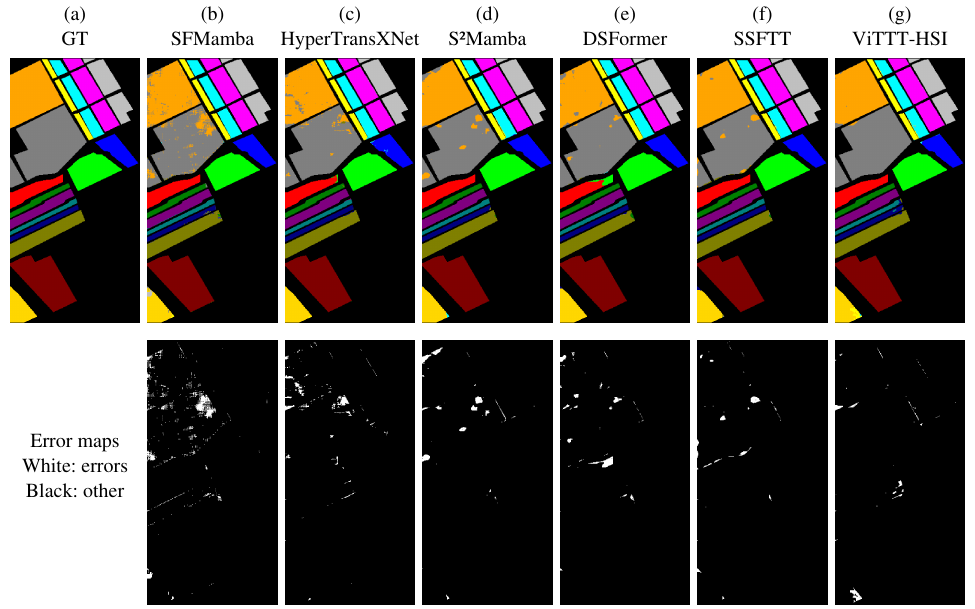}
\caption{Classification maps on Salinas. Panel order and the error-map convention follow Fig.~\ref{fig:map-indian}.}
\label{fig:map-salinas}
\par\vspace{10pt}
\includegraphics[width=0.84\textwidth]{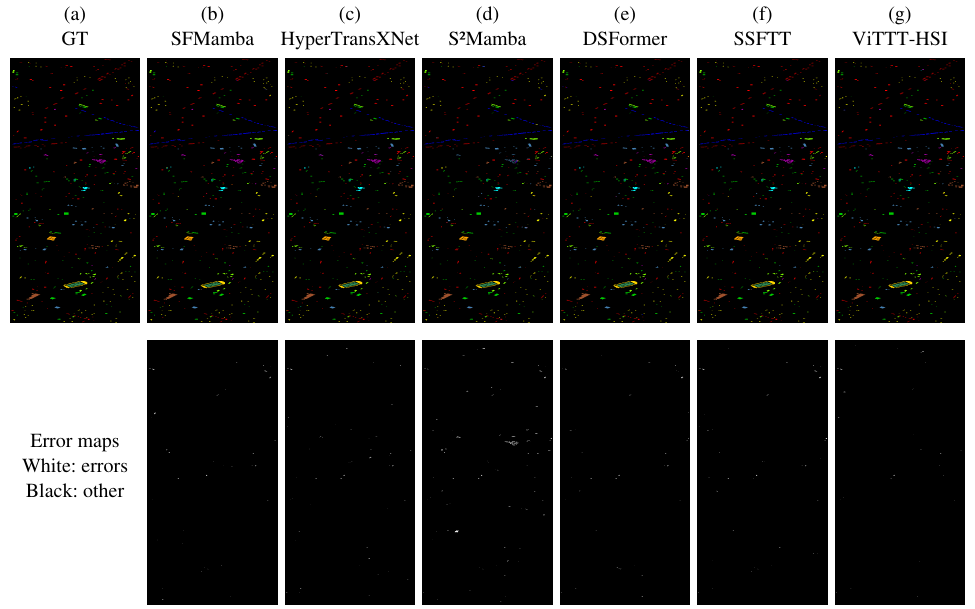}
\caption{Classification maps on Houston 2013. Panel order and the error-map convention follow Fig.~\ref{fig:map-indian}. All maps are rotated $90^\circ$ counterclockwise for readability.}
\label{fig:map-houston13}
\end{figure}

Figures~\ref{fig:map-indian}--\ref{fig:map-houston13} compare six classifiers using consistent colors within each dataset. Predictions are paired with binary error maps (white: incorrect; black: other pixels). ViTTT-HSI uses the validation-selected checkpoint from seed 3405 (Sec.~\ref{sec:checkpoint}).

\FloatBarrier

\subsection{Training dynamics and validation convergence}
\begin{figure}[!htbp]
\centering
\includegraphics[width=0.98\textwidth]{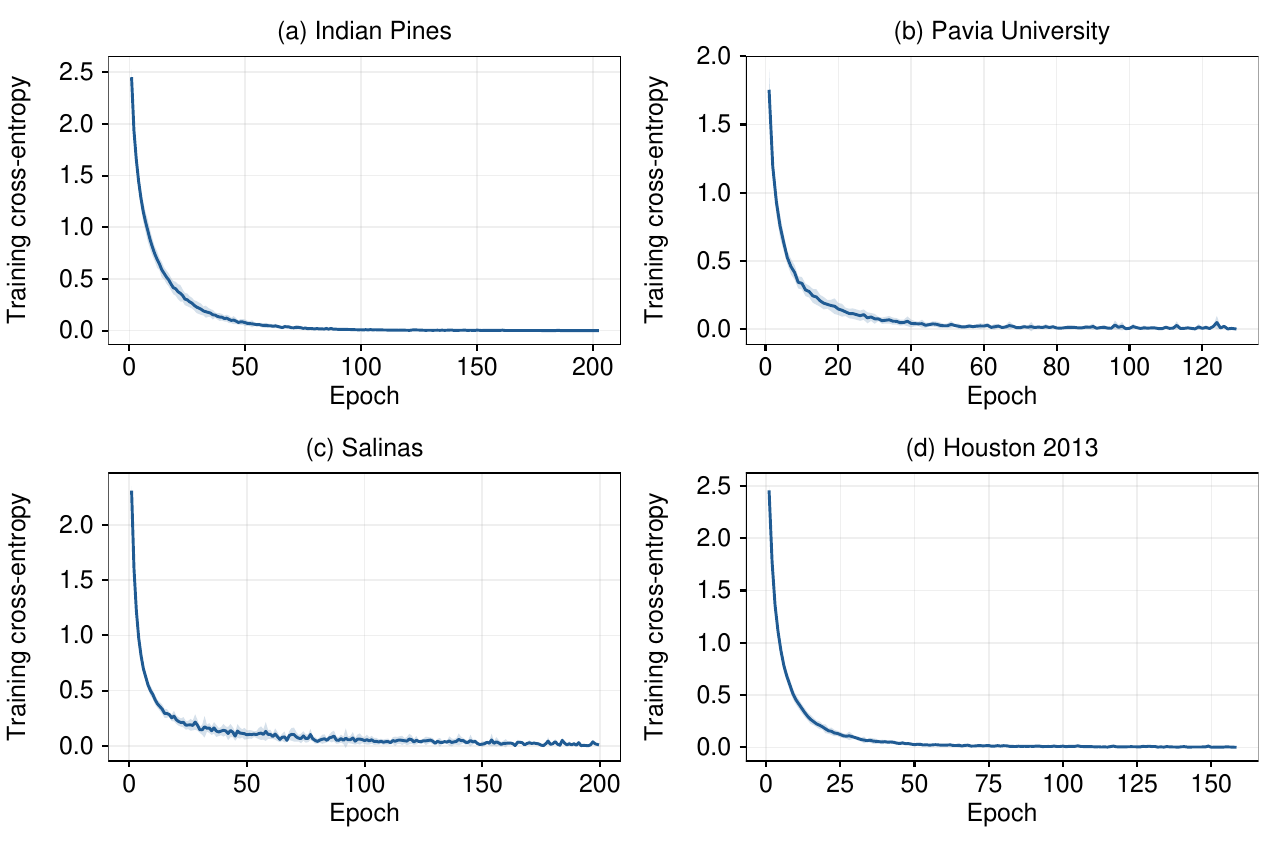}
\caption{Training cross-entropy across seeds 3405--3414: (a) Indian Pines; (b) Pavia University; (c) Salinas; (d) Houston 2013. Lines and bands show the mean and $\pm$ one sample standard deviation of the available runs at each epoch; stopped runs are not extended.}
\label{fig:curve-train_loss}
\end{figure}

\begin{figure}[!htbp]
\centering
\includegraphics[width=0.90\textwidth]{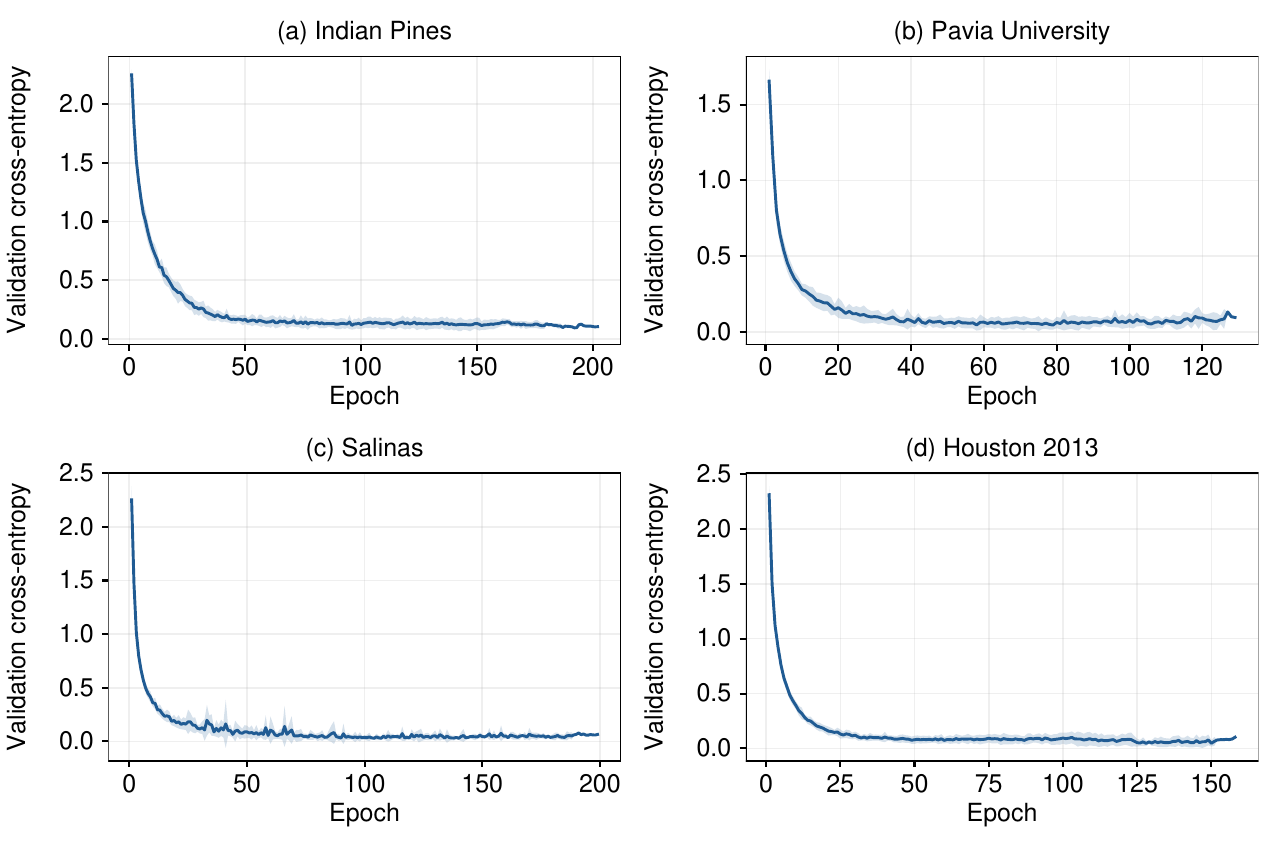}
\caption{Validation cross-entropy across seeds 3405--3414. The line/band convention and panels (a)--(d) follow Fig.~\ref{fig:curve-train_loss}; stopped runs are not extended.}
\label{fig:curve-val_loss}
\par\vspace{10pt}
\includegraphics[width=0.90\textwidth]{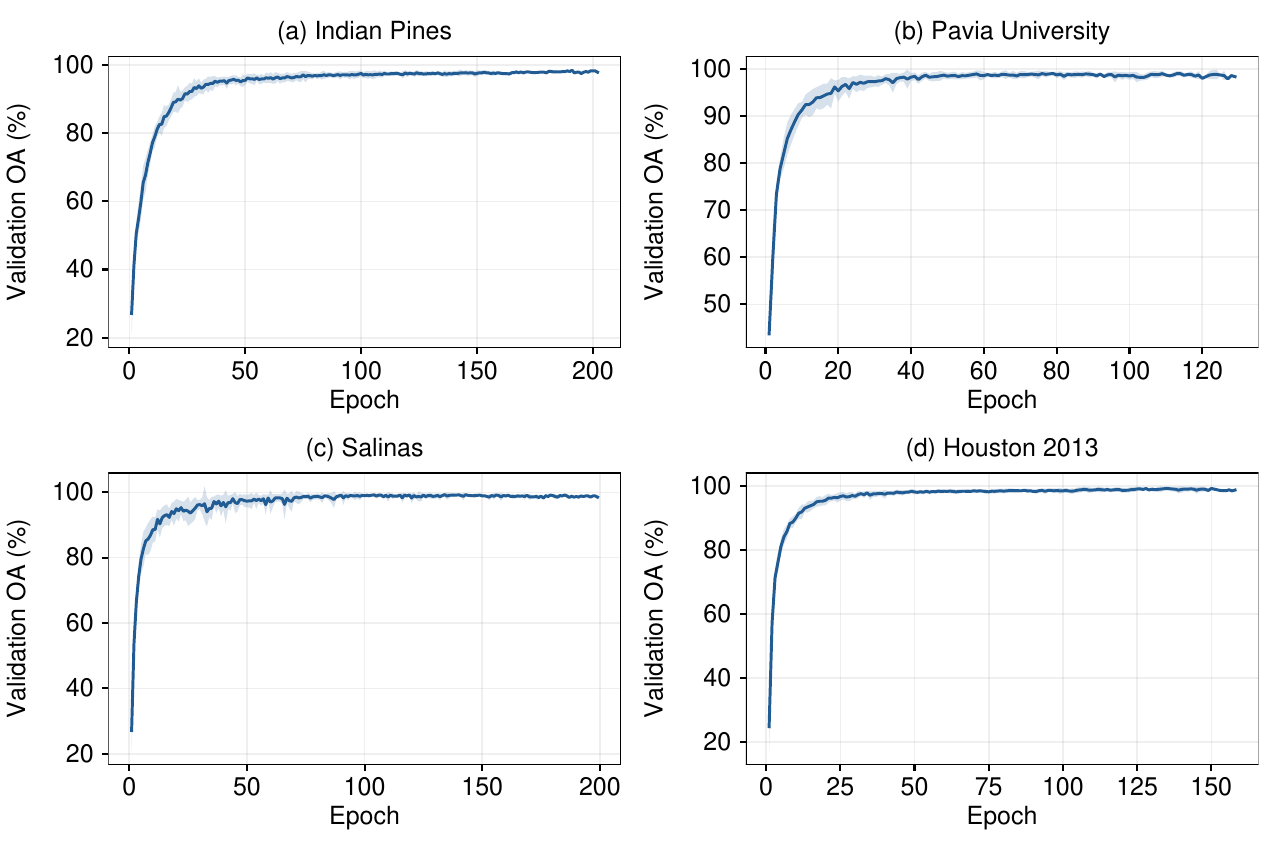}
\caption{Validation overall accuracy across seeds 3405--3414. The line/band convention and panels (a)--(d) follow Fig.~\ref{fig:curve-train_loss}; stopped runs are not extended.}
\label{fig:curve-val_oa}
\end{figure}

The loss and accuracy histories provide a complementary view of optimization under the specified training and checkpoint-selection protocol. Figures~\ref{fig:curve-train_loss}--\ref{fig:curve-val_oa} plot the epoch-wise mean training loss, validation cross-entropy, and validation OA for all four datasets. The curves show rapid early improvements followed by a more gradual refinement of the validation metrics. The two types of validation measurement serve different purposes: OA determines the highest accuracy used in checkpoint selection, while cross-entropy resolves ties between epochs attaining that level. The plotted means use the runs that have reached the respective epoch, because individual runs stop at different epochs with an early-stopping patience of 50 epochs.

\FloatBarrier
\subsection{Effect of spectral embedding and TTT}
Table~\ref{tab:core} provides a factorial ablation of spectral versus linear embedding and the presence of the TTT mixer on Indian Pines and Pavia University. The TTT mixer improves Indian Pines classification with either embedding, and the spectral stem produces the highest OA and AA on Indian Pines. Pavia University also benefits from TTT, while linear embedding with TTT gives a slightly higher mean OA than spectral embedding with TTT. The two datasets therefore exhibit different sensitivities to the embedding pathway.

\begin{table}[tbp]
\caption{Embedding and TTT ablation on Indian Pines and Pavia University (\%, ten seeds).}
\label{tab:core}
\centering\footnotesize
\begin{tabular}{llcc}
\toprule
Dataset & Variant & OA & AA \\
\midrule
Indian & Linear, no TTT & 93.76 $\pm$ 0.88 & 89.45 $\pm$ 3.13 \\
Indian & Spectral, no TTT & 94.19 $\pm$ 0.95 & 90.91 $\pm$ 2.16 \\
Indian & Linear, TTT & 97.42 $\pm$ 0.46 & 94.85 $\pm$ 1.89 \\
Indian & Spectral, TTT & 97.64 $\pm$ 0.58 & 95.18 $\pm$ 1.83 \\
Pavia & Linear, no TTT & 98.42 $\pm$ 0.38 & 97.72 $\pm$ 0.64 \\
Pavia & Spectral, no TTT & 98.46 $\pm$ 0.35 & 97.78 $\pm$ 0.69 \\
Pavia & Linear, TTT & 98.97 $\pm$ 0.35 & 98.67 $\pm$ 0.56 \\
Pavia & Spectral, TTT & 98.88 $\pm$ 0.49 & 98.51 $\pm$ 0.72 \\
\bottomrule
\end{tabular}
\end{table}

\subsection{Role of the TTT inner update}
Table~\ref{tab:inner} compares the full mixer against a model without the mixer and a frozen-inner version that retains the inner-model architecture but skips fast-weight updates. Full TTT improves Indian Pines OA by 3.5555 percentage points over the frozen-inner version, with a smaller gain of 0.2478 percentage points on Pavia University. These controlled comparisons isolate the benefit of patch-specific fast-weight updates more clearly than removing the mixer altogether: the learned inner-model architecture is retained in the frozen variant, whereas only its data-conditioned update is omitted.

\begin{table}[tbp]
\caption{Ablation of the TTT inner update (\%, mean $\pm$ standard deviation; ten seeds).}
\label{tab:inner}
\centering\footnotesize
\begin{tabular}{llcc}
\toprule
Dataset & Mixer setting & OA & AA \\
\midrule
Indian & No TTT & 94.19 $\pm$ 0.95 & 90.91 $\pm$ 2.16 \\
Indian & Frozen inner & 94.08 $\pm$ 0.65 & 90.18 $\pm$ 2.06 \\
Indian & Full & 97.64 $\pm$ 0.58 & 95.18 $\pm$ 1.83 \\
Pavia & No TTT & 98.46 $\pm$ 0.35 & 97.78 $\pm$ 0.69 \\
Pavia & Frozen inner & 98.63 $\pm$ 0.31 & 98.11 $\pm$ 0.52 \\
Pavia & Full & 98.88 $\pm$ 0.49 & 98.51 $\pm$ 0.72 \\
\bottomrule
\end{tabular}
\end{table}

\subsection{Center--global features and patch-size analysis}
The center--global classifier is assessed against center-only and global-only heads in Table~\ref{tab:head}. The center feature is especially important: global-only classification is markedly weaker on both datasets. Combining center and global features improves Indian Pines OA over center-only classification, while the center-only head is slightly better on Pavia University. The result is consistent with the different contributions of center-pixel evidence and neighborhood context across scenes.

\begin{table}[tbp]
\caption{Classification-head ablation (\%, ten-seed OA mean $\pm$ standard deviation).}
\label{tab:head}
\centering\footnotesize
\begin{tabular}{llc}
\toprule
Dataset & Head & OA \\
\midrule
Indian & Center & 97.43 $\pm$ 0.40 \\
Indian & Global & 95.04 $\pm$ 0.69 \\
Indian & Center + Global & 97.64 $\pm$ 0.58 \\
Pavia & Center & 98.90 $\pm$ 0.41 \\
Pavia & Global & 94.71 $\pm$ 0.71 \\
Pavia & Center + Global & 98.88 $\pm$ 0.49 \\
\bottomrule
\end{tabular}
\end{table}

Table~\ref{tab:patch} evaluates patch sizes 11, 21, and 29 under both full TTT and a no-TTT setting, using three seeds per configuration. The full TTT configuration obtains higher mean OA than the no-TTT setting at each evaluated patch size. The observed gain is largest at $P=29$ on Indian Pines and $P=21$ on Pavia University, indicating that the interaction between patch context and inner adaptation depends on the dataset.

\begin{table}[tbp]
\caption{Patch-size ablation on Indian Pines and Pavia University (\%, mean OA over three seeds). The last column is the Full TTT minus No TTT difference in percentage points.}
\label{tab:patch}
\centering\footnotesize
\begin{tabular}{lrrrr}
\toprule
Dataset & $P$ & Full TTT & No TTT & Difference \\
\midrule
Indian & 11 & 97.34 & 95.30 & +2.04 \\
Indian & 21 & 97.24 & 95.17 & +2.07 \\
Indian & 29 & 97.69 & 94.80 & +2.89 \\
Pavia & 11 & 97.46 & 97.36 & +0.11 \\
Pavia & 21 & 98.74 & 98.21 & +0.53 \\
Pavia & 29 & 98.72 & 98.50 & +0.22 \\
\bottomrule
\end{tabular}
\end{table}

\subsection{Spatially separated evaluation}
A further evaluation on Pavia University uses a spatial-block split with patch size 11 and three seeds. As summarized in Table~\ref{tab:spatial}, full TTT outperforms the no-TTT variant in this more spatially separated setup. The full TTT configuration achieves 90.16\% OA, exceeding the no-TTT configuration by 1.40 percentage points in this spatial-block experiment.

\begin{table}[tbp]
\caption{Spatial-block evaluation on Pavia University using patch size 11 (\%, mean $\pm$ standard deviation across three seeds).}
\label{tab:spatial}
\centering\footnotesize
\begin{tabular}{lcc}
\toprule
Variant & OA & AA \\
\midrule
Full & 90.16 $\pm$ 1.04 & 87.39 $\pm$ 2.65 \\
No TTT & 88.76 $\pm$ 0.85 & 85.56 $\pm$ 6.92 \\
\bottomrule
\end{tabular}
\end{table}

\subsection{Computational complexity and efficiency}
Table~\ref{tab:eff} presents the parameter count, computational cost, and training and inference times of ViTTT-HSI on an NVIDIA RTX 3090. Computational cost in billions of floating-point operations (GFLOPs) is estimated for one input patch with a batch size of one, counting the main forward computations of both TTT inner adaptation and query evaluation at two floating-point operations per multiply--accumulate. Training time covers the epoch loop, including validation and checkpoint writing. Testing time is measured on the entire test loader after warm-up, with device synchronization.

\begin{table}[tbp]
\caption{Model complexity and seed-3405 runtime of ViTTT-HSI. GFLOPs denote the estimated major forward arithmetic operations per input patch; M denotes millions of parameters, and runtime is reported in seconds.}
\label{tab:eff}
\centering\footnotesize
\begin{tabular}{lrrrr}
\toprule
Dataset & Parameters (M) & GFLOPs & Train (s) & Test (s) \\
\midrule
Indian & 0.0923 & 0.167 & 139.1 & 4.04 \\
Pavia & 0.0914 & 0.165 & 42.2 & 11.15 \\
Salinas & 0.0923 & 0.145 & 93.3 & 20.44 \\
Houston 2013 & 0.0922 & 0.166 & 219.5 & 4.33 \\
\bottomrule
\end{tabular}
\end{table}

Across all four datasets, the model uses approximately 0.09 million trainable parameters. The measured test-set runtime varies with the number of pixels evaluated and the dataset-specific patch size.

\FloatBarrier
\section{Conclusion}
\label{sec:conclusion}
This study adapted the ViT$^3$ visual test-time training mechanism to center-pixel hyperspectral classification through a spectral-aware embedding, spatial-token-preserving TTT encoders, and a center--global classifier. Four public benchmarks and ten random seeds supported competitive classification performance under the specified low-label random-pixel protocols. Controlled ablations indicate that the patch-conditioned TTT inner update contributes to the classification gains, especially on Indian Pines, while the benefits of the spectral stem and global feature pooling are dependent on the dataset. The results establish an experimental basis for applying patch-conditioned visual TTT to spectral--spatial center-pixel classification.

\subsection*{Disclosures}
The authors declare that there are no financial interests, commercial affiliations, or other potential conflicts of interest that could have influenced the objectivity of this research or the writing of this paper.

\subsection*{Code and Data Availability}
The Indian Pines, Pavia University, Salinas, and Houston 2013 datasets used in this study are available through the HSI Datasets repository on Hugging Face~\cite{hsiDatasetsRepository}. The source code and reproduction scripts for ViTTT-HSI will be released in the \href{https://github.com/yangshidi/Spectral-spatial-vision-test-time-training-for-hyperspectral-image-classification}{project GitHub repository}.

\subsection*{Acknowledgments}
We thank OpenAI ChatGPT for assistance with English-language editing and LaTeX preparation. The authors take full responsibility for the scientific content, accuracy, and conclusions of this manuscript.

\begin{spacing}{1}
\raggedright
\urlstyle{same}
\bibliography{references}
\bibliographystyle{spiejour}
\end{spacing}

\vspace{2ex}
\noindent\textbf{Xiang Zhao} received his B.Sc. degree in mathematics
and applied mathematics from Harbin Engineering University, Harbin,
China, in 2025. He is currently pursuing an M.Sc. degree in mathematics
at the College of Mathematical Sciences, Harbin Engineering University.
His research interests include remote sensing image processing,
hyperspectral image classification, and deep learning algorithms,
particularly graph neural networks and attention mechanisms.

\end{spacing}
\end{document}